\documentclass[twocolumn]{article}

\usepackage{amsmath}
\usepackage{amssymb}
\usepackage{amsfonts}
\usepackage{array}
\usepackage[usenames]{color}
\usepackage{wrapfig}
\usepackage[margin=1in]{geometry}
\usepackage{graphicx}
\usepackage{epstopdf}
\usepackage{caption}
\usepackage{subcaption}
\usepackage{hyperref}

\newcommand{\blockpar}{\parindent 0pt \parskip 5pt}

\blockpar

\newcommand{\tup}[1]{\left< #1 \right>}			

\newcommand{\func}[3]{#1 \colon #2 \rightarrow #3}	
\newcommand{\fig}[1]{Fig.~\ref{#1}}
\newcommand{\rem}[1]{}

\newcommand{\st}{ \mathrel{\colon} }	

\newcommand{\code}[1]{{\texttt{#1}}}

\newcommand{\sub}{\subseteq}		
\newcommand{\define}{ \mathrel{\bf \colon \kern -2pt =} }	
\newcommand{\ints}{\mathbb{Z}}

\newcommand{\hg}{{\cal H}}
\newcommand{\graph}{{\cal G}}

\newcommand{\hpair}[2]{#1 \!\colon\!\! #2}
\newcommand{\adj}{\hbox{adj}}
\newcommand{\inc}{\hbox{inc}}

\usepackage{scalerel,stackengine}
\stackMath
\newcommand\widecheck[1]{%
\savestack{\tmpbox}{\stretchto{%
  \scaleto{%
    \scalerel*[\widthof{\ensuremath{#1}}]{\kern-.6pt\bigwedge\kern-.6pt}%
    {\rule[-\textheight/2]{1ex}{\textheight}}
  }{\textheight}%
}{0.5ex}}%
\stackon[1pt]{#1}{\scalebox{-1}{\tmpbox}}%
}

\newcommand{\red}{\textcolor{red}}

\newcommand{\mypng}[3]{
	\begin{figure}[htbp]
	\begin{center}
	\includegraphics[scale=#1]{#2.png}
	\caption{\small #3}
	\label{#2}
	\end{center}
	\end{figure}
}

\newcommand{\kskel}{k\mbox{-skel}}
\newcommand{\vem}{\vspace{-1em}}
\newcommand{\vhalf}{\vspace{-.5em}}

\begin{document}

\title{{\bf Hypernetwork Science: \\ From Multidimensional Networks to \\ Computational Topology}\thanks{PNNL PNNL-SA-152208}
}

\author{
	Cliff A.\ Joslyn\thanks{Pacific Northwest National Laboratory, Seattle, WA}, 
	Sinan Aksoy\thanks{Pacific Northwest National Laboratory, Richland, WA}, 
	Tiffany J.\ Callahan\thanks{U.\ of Colorado Anschutz Medical Campus, Denver, CO},
	Lawrence E.\ Hunter\footnotemark[4],	\\
	Brett Jefferson\footnotemark[3],	
	Brenda Praggastis\footnotemark[2], 
	Emilie A.H.\ Purvine\footnotemark[2],
	Ignacio J.\ Tripodi\footnotemark[4]
}

\date{March, 2020}

\maketitle

\begin{abstract}

As data structures and mathematical objects used for complex systems modeling, hypergraphs sit nicely poised between on the one hand the world of network models, and on the other that of higher-order mathematical abstractions from algebra, lattice theory, and topology. They are able to represent complex systems interactions more faithfully than graphs and networks, while also being some of the simplest classes of systems representing topological structures as collections of  multidimensional objects connected in a particular pattern. In this paper we discuss the role of (undirected) hypergraphs in the science of complex networks, and provide a mathematical overview of the core concepts needed for hypernetwork modeling, including duality and the relationship to bicolored graphs, quantitative adjacency and incidence, the nature of walks in hypergraphs, and available topological relationships and properties. We close with a brief discussion of two example applications: biomedical databases for disease analysis, and domain-name system (DNS) analysis of cyber data.

\end{abstract}

\rem{

Hypergraph mathematics
	Duality
	swalks
	Homology
	Bipartites
Hypergraph analytics
	Collapsing
	Density
	Inclusivity

Ask for
	DNS homologies
	CU data writeup

}

\vem

\section{Hypergraphs for Complex Systems Modeling}

In the study of complex systems, graph theory has been the mathematical scaffold underlying network science  \cite{Barabasi2016}. A graph $\graph = \tup{V, E}$ comprises a set  $V$ of vertices connected in a set $E \subseteq {V \choose 2}$ of edges (where ${V \choose 2}$ here means all unordered pairs of $v \in V$), where each edge $e \in E$ is a pair of distinct vertices. Systems studied in biology, sociology, telecommunications, and physical infrastructure often afford a representation as such a set of entities with binary relationships, and hence may be analyzed utilizing graph theoretical methods. 

Graph models benefit from simplicity and a degree of universality. But as abstract mathematical objects, graphs are limited to representing {\it pairwise} relationships between entities, while real-world phenomena in these systems can be rich in {\em multi-way} relationships involving interactions among more than two entities, dependencies between more than two variables, or properties of collections of more than two objects. 
Representing {\em group} interactions is not possible  in graphs natively, but rather requires either more complex mathematical objects, or coding schemes like ``reification'' or  semantic labeling in bipartite graphs.
Lacking
multi-dimensional relations, it is hard to address questions of
``community interaction'' in graphs: e.g.,\ how is a collection of
entities $A$ connected to another collection $B$ through chains of
other communities?; where does a particular community stand in
relation to other communities in its neighborhood?

The mathematical object which {\em natively} represents multi-way interactions in networks is called a  ``hypergraph''  \cite{berge1973graphs}.%
	\footnote{Throughout this paper we will deal only with ``basic'' hypergraphs in the sense of being undirected, unordered, and unlabeled. All of these forms are available and important \cite{AuGFaP01,GaGLoG93}.}
	 In contrast to a graph, in a hypergraph $\hg = \tup{V, E}$ those same vertices are now connected generally in a family $E$ of hyperedges, where now a hyperedge $e \in E$ is an arbitrary subset $e \sub V$ of $k$ vertices, thereby representing a $k$-way relationship for any integer $k > 0$.
Hypergraphs are thus the natural representation of a broad range of systems, including those with the kinds of multi-way relationships mentioned above.
Indeed, hypergraph-structured data (i.e.\ hypernetworks) are ubiquitous, occurring whenever information presents naturally as set-valued, tabular, or bipartite data. 

\begin{figure}[h]
\centering
\begin{subfigure}[c]{0.17\textwidth}
    \centering
    \includegraphics[width=\textwidth]{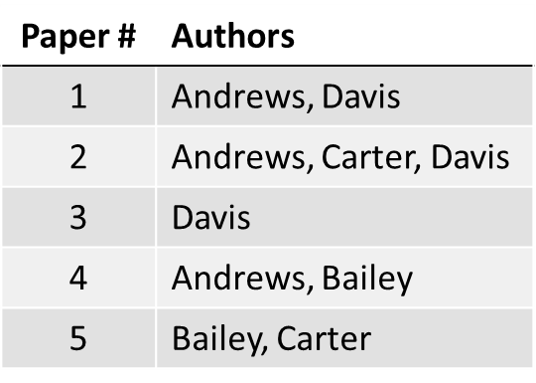}
    \label{fig:biblio_list}
\end{subfigure}
~
\begin{subfigure}[c]{0.22\textwidth}
    \centering
    \includegraphics[width=\textwidth]{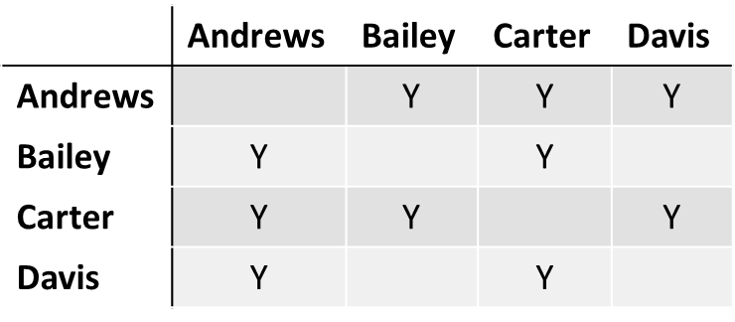}
    \label{fig:biblio_adj}
\end{subfigure}
\begin{subfigure}[c]{0.22\textwidth}
    \centering
    \includegraphics[width=\textwidth]{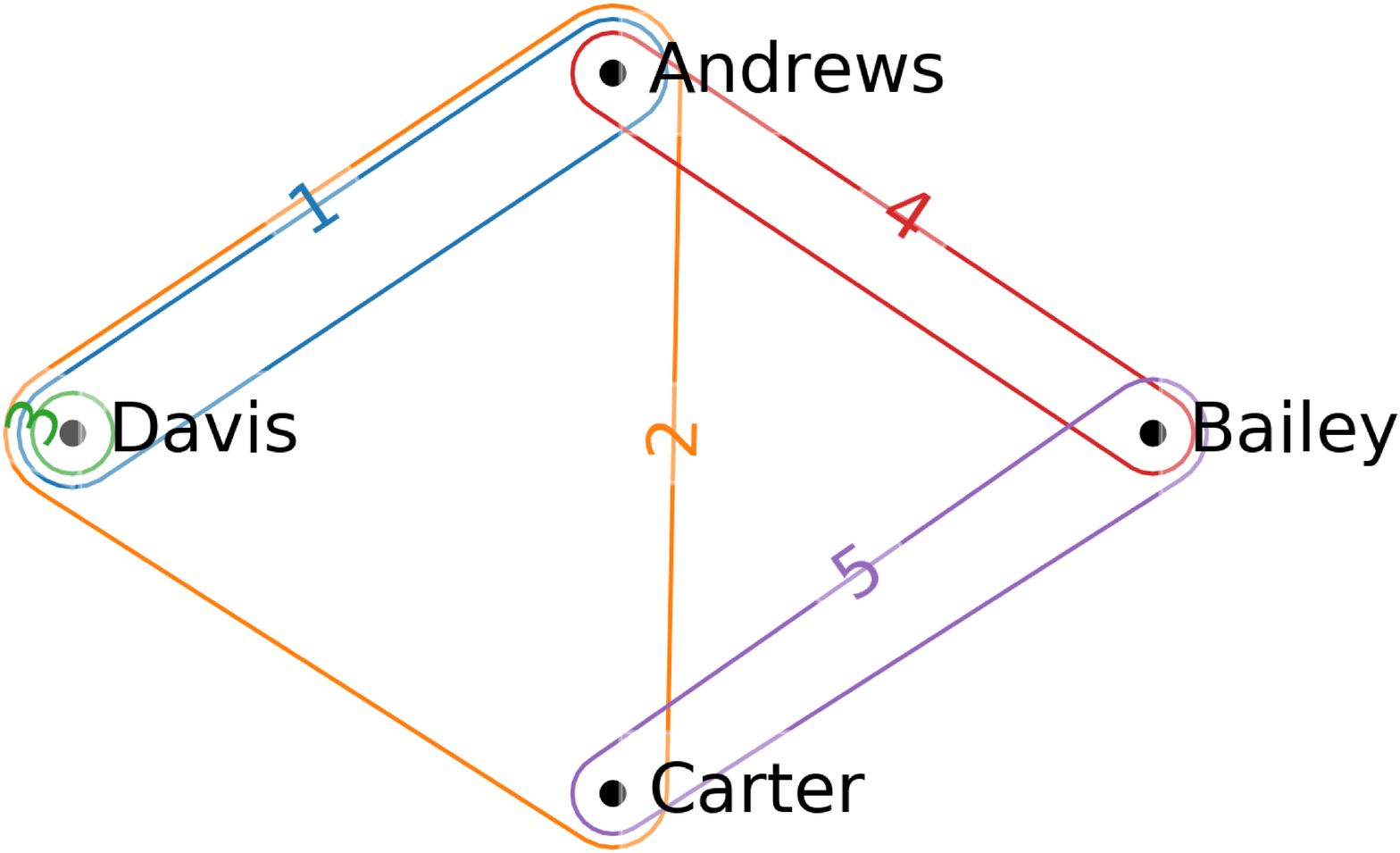}
    \label{fig:biblio_hg}
\end{subfigure}
\begin{subfigure}[c]{0.22\textwidth}
    \centering
    \includegraphics[width=\textwidth]{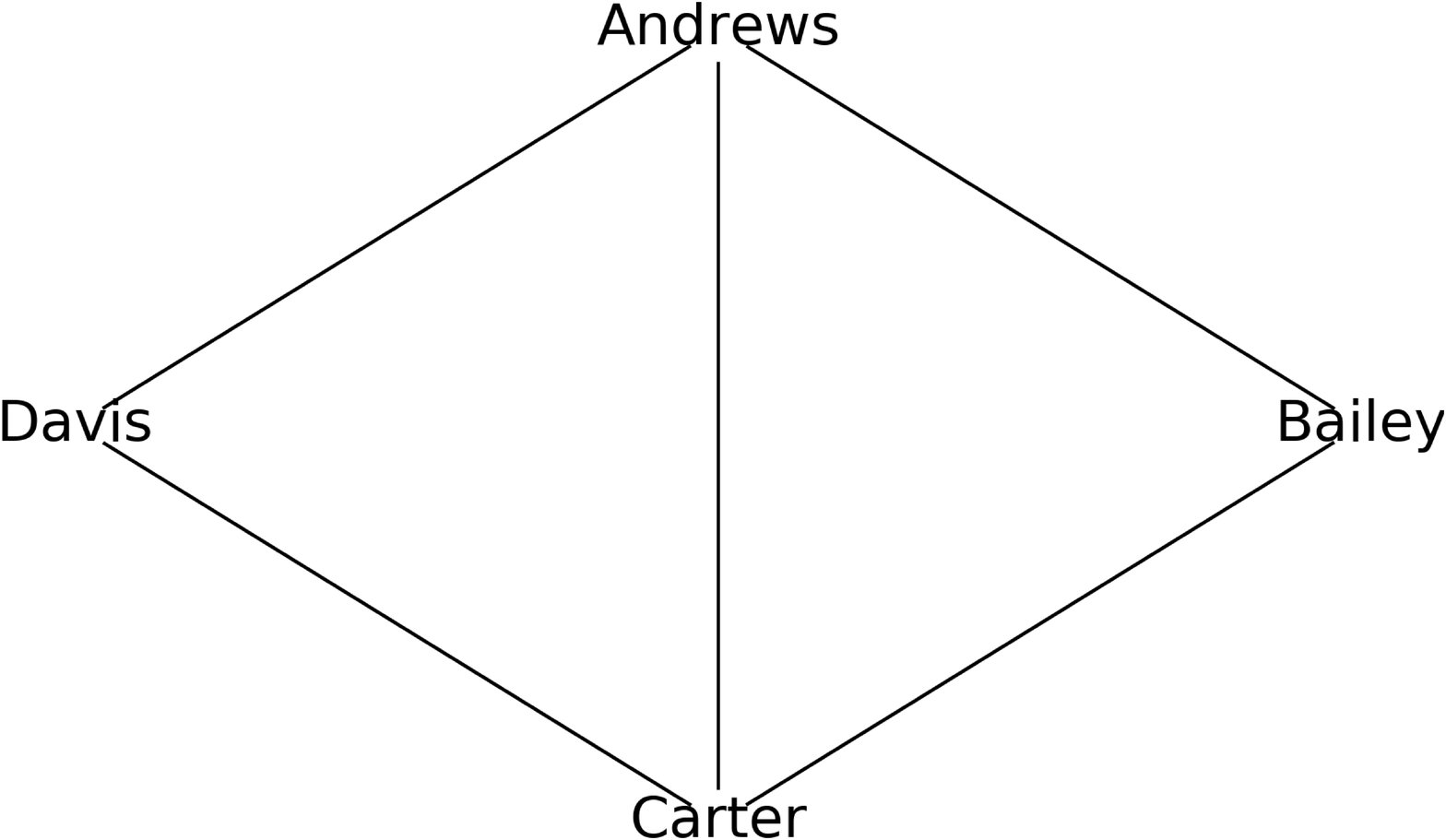}
    \label{fig:biblio_twosec}
\end{subfigure}
\vem
\caption{\small Bibliometrics example comparing graphs and hypergraphs. (Upper left) Collaborative authorship structure of a set of papers. (Lower left) Euler diagram of its hypergraph. (Upper right) Adjacency matrix of the 2-section. (Lower right) Co-authorship graph resulting from the 2-section.}
\label{fig:hgraph}
\end{figure}

Hypergraph models are definitely more complicated than graphs, but the price paid allows for higher fidelity representation of data which may contain multi-way relationships. An example from bibliometrics is shown in Figure~\ref{fig:hgraph}.%
		   \footnote{Hypergraph calculations shown in this paper were produced using PNNL's open source hypergraph analytical capabilities HyperNetX (HNX, {\tt https://github.com/pnnl/HyperNetX}) and the Chapel Hypergraph Library (CHGL, {\tt https://github.com/pnnl/chgl}); and additionally diagrams were produced in HNX.}
		   On the upper left is a table showing a selection of five papers co-authored by different collections of four authors. Its hypergraph is shown in the lower left in the form of an ``Euler diagram'', with hyperedges as colored bounds around groups of vertices. A typical approach to these same data would be to reduce the collaborative structure to its so-called ``2-section'': a graph of all and only the two-way interactions present, whether explicitly listed (like paper 1) or implied in virtue of larger collaborations (as for paper 2). That co-authorship graph is shown in the lower right, represented by its  adjacency matrix  in the upper right. It can be seen that the reduced graph form necessarily loses a great deal of information, ignoring information about the single-authored paper (3) and the three-authored paper (2).

Hypergraphs are
closely related to important  objects in discrete mathematics used in data science such as bipartite
graphs, set systems, partial orders, finite topologies, simplicial complexes, and especially
graphs proper, which they explicitly generalize: every graph is a
2-uniform hyergraph, so that $|e|=k=2$ for all hyperedges $e \in E$.
Thus they support a wider range of mathematical methods, such as those from computational topology, to identify features specific to the high-dimensional complexity in hypernetworks, but not available using graphs. 

Hypergraph methods are well known in discrete mathematics and computer science, 
where, for example, hypergraph partitioning methods help enable parallel matrix
computations \cite{Devine2006}, and have
applications in VLSI \cite{Karypis2000}.  In the
network science literature, researchers have devised several path and
motif-based hypergraph data analytics (albeit fewer than their graph
counterparts), such as in clustering coefficients
\cite{Robins2004} and
centrality metrics \cite{Estrada2006}.
Although an expanding body of research attests to the increased utility of hypergraph-based analyses \cite{IaIPeG19,JoJ13,klamt2009hypergraphs,PaAPeG17}, and are seeing increasingly wide adoption \cite{JaMLuL18,LeWReG19,MiM00}, 
many network science methods have been historically developed explicitly (and often, exclusively) for graph-based analyses. Moreover, it is common for analysts to reduce data arising from hypernetworks to graphs, thereby losing critical information.

As explicit generalizations of graphs, we must take care with axiomatization, as there are many, sometimes mutually inconsistent, sets of possible definitions of hypergraph concepts which can yield the same results (all consistent with graph theory) when instantiated to the 2-uniform case.
And some graph concepts have difficulty extending naturally at all. 
For example, extending the spectral theory of graph adjacency matrices to hypergraphs is unclear:  hyperedges may contain more than two vertices, so the usual (two-dimensional) adjacency matrix cannot code adjacency relations. 
In other cases, hypergraph extensions of graph theoretical concepts may be natural, but trivial, and risk ignoring structural properties only in hypergraphs. 
For example, while edge incidence and vertex adjacency can occur in at most one vertex or edge for graphs, these notions are set-valued and hence {\it quantitative} for hypergraphs.  
So while subsequent graph walk concepts like connectivity are applicable to hypergraphs, if applied simply, they ignore high-order structure in not accounting for the ``widths'' of hypergraph walks. 

Researchers have handled the complexity and ambiguity of hypergraphs in different ways. 
A very common approach is to limit attention to $k$-uniform hypergraphs, where all edges have precisely $k$ vertices (indeed, one can consider graph theory itself as actually the theory of $2$-uniform hypergraphs). This is the case with 
much of the hypergraph mathematics literature, including hypergraph coloring \cite{dinur2005hardness,krivelevich2003approximate}, hypergraph spectral theory \cite{chung1993laplacian,cooper2012spectra}, hypergraph transversals \cite{alon1990transversal}, and extremal problems \cite{rodl2004regularity}.
$k$-uniformity is a very strong assumption, which supports the identification of mathematical results. But real-world hypergraph data are effectively {\em never} uniform; or rather, if presented with uniform hypergraph data, a wise data scientists would be led to consider other mathematical representations for parsimony. 

Another prominent approach to handling real-world, and thus non-uniform, hypergraph data is to study simpler graph structures implied by a particular hypergraph. Known by many names, including
line graph, 2-section, clique expansion, and one-mode projection, such reductions allow application of standard graph-theoretic tools.
Yet, unsurprisingly, such hypergraph-to-graph reductions are inevitably and strikingly lossy \cite{Dewar2018, Kirkland2017}. 
Hence, although affording simplicity, such approaches are of limited utility in uncovering hypergraph structure.

Our research group is dedicated to facing the challenge of the
complexity of hypergraphs in order to gain the formal clarity and support for
analysis of complex data they provide. We recognize that to enable analyses of hypernetwork data to better reflect their complexity but remain tractable and applicable,  striking a balance in this faithfulness-simplicity tradeoff is essential. Placing hypergraph methods in the context of the range of both network science methods on the one hand, and higher-order topological methods on the other, can point the way to such a synthesis.

The purpose of this paper is to communicate the {\em breadth} of hypergraph methods (which we are exploring in depth elsewhere) to the complex systems community. In the next section we survey the range of mathematical methods and data structures we use. Following that we will illustrate some uses by showing examples in two different contexts:  gene set annotations for disease analysis and drug discovery, and cyber analytics of domain-name system relations.

\rem{Critical to this approach for us is to use {\it high-order hypergraph walks} or ``$s$-walks'' \cite{AkSJoC19}, where the order $s$ controls the minimum walk ``width" in terms of edge overlap size. 
High-order $s$-walks ($s>1$) are possible on hypergraphs whereas for graphs, all walks are 1-walks. The hypergraph walk-based methods we develop include connected component analyses, graph-distance based metrics such as closeness-centrality, and motif-based measures such as clustering coefficients. }

\vem

\section{Hypergraph Analytics}

\rem{

We now introduce the mathematical foundatinos of hypergraph modeling and methodology. In summary, natively, graphs represent {\em uniformly} the {\em pairwise} interactions of the values of {\em one} variable; while hypergraphs natively represent {\em non-uniformly} the {\em multiway} interactions of the values of {\em two} variables. 

}

A {\bf hypergraph} is a structure $\hg = \tup{V,E}$, with $V=\{v_j\}_{j=1}^n$  a  set of vertices, and $E = \{e_i\}_{i=1}^m$ an indexable family of hyperedges $e_i \sub V$. Hyperedges  come in different sizes $|e_i|$ possibly ranging from the singleton $\{v\} \sub V$ (distinct from the element $v \in V$) to the entire vertex set $V$.

A hyperedge $e=\{u,v\}$ with $|e|=2$ is the same as a graph edge. Indeed, all graphs ${\cal G} = \tup{V,E}$  are hypergraphs: in particular, graphs are ``2-uniform'' hypergraphs, so that now $E \sub {V \choose 2}$ and all $e \in E$ are unordered pairs with $|e|=2$. It follows that concepts and methods in hypergraph theory should explicitly specialize to those in graph theory for the 2-uniform case. But conversely, starting from {\em graph theory} concepts, there can be many ways of consistently extending them to hypergraphs. The reality of this will be seen in a number of instances below.

Hypergraphs can be represented in many forms. In our example in \fig{fig:hgraph} above, letting $V = \{ a,b,c,d \}$ for the authors and $E = \{ 1, 2, 3, 4, 5 \}$ for the papers, we first represent it as a set system
	\[ \hg = \{ \{ a, d \}, \{ a, c, d \}, \{ d \}, \{ a, b \}, \{ b, c \} \}.	\]
We commonly compress set notation for convenience, and when including edge names as well, this yields the compact set system form $\hg = \{ \hpair{1}{ad}, \hpair{2}{acd}, \hpair{3}{d}, \hpair{4}{ab}, \hpair{5}{bc} \}$. This representation in turn points to the fact that a hypergraph $\hg$ is  determined uniquely by its Boolean {\bf incidence matrix} $B_{n \times m}$, where $B_{ji}=1$ iff $v_j \in e_i$, and 0 otherwise. The incidence matrix for the example from Figure~\ref{fig:hgraph} is shown in Table~\ref{tab:hgraph}. Note that hypergraph incidence matrices are general Boolean matrices, rectangular and non-symmetric, unlike the adjacency matrices typically used to define graphs; while the incidence matrices of graphs are restricted to having precisely two 1's in each column. Also, adjacency structures for hypergraphs are substantially more complicated than for graphs, and in fact are not matrices at all.

\begin{table}[h]
	\footnotesize
    \centering
    \begin{tabular}{l|ccccc}
                & 1 &   2   &   3   &   4   &   5   \\
                \hline
         $a$ &  1 & 1 & 0 & 1 & 0 \\
         $b$ & 0 & 0 & 0 & 1 & 1\\
         $c$ & 0 & 1 & 0 & 0 & 1 \\
         $d$ & 1 & 1 & 1 & 0 & 0
    \end{tabular}
    \caption{\small Incidence matrix of example hypergraph \fig{fig:hgraph}.}
    \label{tab:hgraph}
\end{table}

The {\bf dual hypergraph}  $\hg^*=\tup{E^*,V^*}$ of $\hg$ has vertex set $E^*=\{e_i^*\}_{i=1}^m$
and family of hyperedges $V^*=\{v_j^*\}_{j=1}^n$, where $v_j^* \define \{e_i^* : v_j \in e_i\}$.   $\hg^*$ is just the hypergraph with the transposed incidence matrix $B^T$, and $(\hg^*)^*=\hg$. 
We thus consider that hypergraphs always present as dual {\em pairs}, which we call the ``forward'' and the ``dual'' somewhat arbitrarily, depending on how the data are naturally presented. But this is not true for graphs: the dual $\graph^*$ of a graph $\graph$ is 2-uniform (and hence still a graph) if and only if $\graph$ is 2-regular (all vertices have degree 2), in which case $\graph$ is a cycle or disjoint union of cycles. The dual of our example is shown in \fig{dual}.

\mypng{.25}{dual}{Dual hypergraph $\hg^*$ of our example.}

\begin{figure}[htbp]
\centering
\begin{subfigure}[c]{0.1\textwidth}
    \centering
    \includegraphics[width=\textwidth]{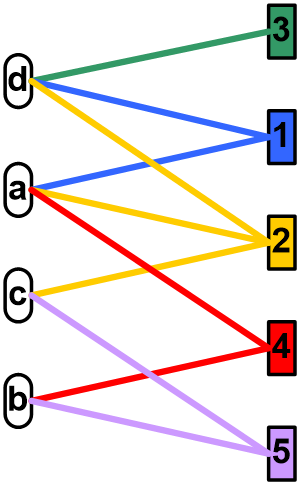}
    \label{fig:biblio_list}
\end{subfigure}
~
\begin{subfigure}[c]{0.1\textwidth}
    \centering
    \includegraphics[width=\textwidth]{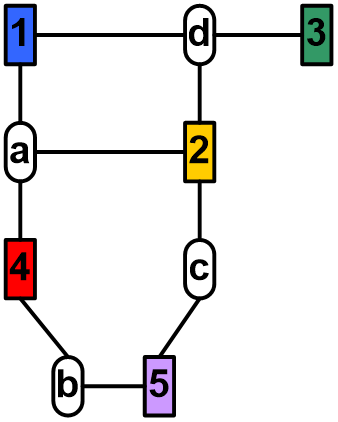}
    \label{fig:biblio_adj}
\end{subfigure}
\caption{\small Two different layouts of the bicolored graph representation of the hypergraph $\hg$.}
\label{fig:bicolored}
\end{figure}

There is a bijection between the class of hypergraphs and that of {\bf bicolored graphs}%
	  \footnote{Typically the concept of a {\bf bipartite} graph is used here, which is a graph that admits to at least one bicoloring function. The resulting differences are interesting, but not significant for this paper.} 
	  ${\cal G} = \tup{ V,E,\phi}$, where now $E \sub {V \choose 2}$ is a set of unordered pairs of vertices (graph edges), and $\func{\phi}{V}{\{0,1\}}$ with $\{v_i,v_j\} \in E$ iff $\phi(v_i) \neq \phi(v_j)$. Conversely, any bicolored graph ${\cal G}$ determines a hypergraph ${\cal H}$ by associating the vertices and hyperedges of $\hg$ with the two colors respectively, and then defining $B_{j,i}=1$ if and only if $\{v_i,v_j\}\in E$.
	  \fig{fig:bicolored} shows two different layouts of the bicolored graph form of our example hypergraph $\hg$.

	  Moving from the bijection between bicolored graphs and hypergraphs to establishing canonical isomorphisms still requires careful consideration, ideally in a categorical context \cite{Dorfler1980, Fong2018, Schmidt2019}. While a number of complex network analytics for bipartite graph data can be applied naturally to hypergraphs, and {\it vice versa}, depending on the semantics of the data being modeled, questions and methods for data with this common structure may be better addressed in one or another form, if only for algorithmic or cognitive reasons. Nor does it mean that graph theoretic methods suffice for studying hypergraphs. Whether interpreted as bicolored graphs or hypergraphs, data with this structure often require entirely different network science methods than (general) graphs. An obvious example is triadic measures like the graph clustering coefficient: these cannot be applied to bicolored graphs since (by definition) bicolored graphs have no triangles. Detailed work developing bipartite analogs of modularity \cite{Barber2007}, community structure inference techniques \cite{Larremore2014}, and other graph-based network science topics \cite{Latapy2008} further attests that bipartite graphs (and hypergraphs) require a different network science toolset than for graphs. 

\rem{

In this paper, we utilize the language of hypergraphs, and not bibcolored graphs, because of the fundamentally set-theoretic nature of our approach. Our focus in this work is on hyperedge incidences and hyperwalks that arise from sequences of incident hyperedges. Hyperedges themselves are defined explicitly for hypergraphs, but only implicitly for bicolored graphs (as the neighborhood of a vertex in the color class designated for hyperedges). 
For this reason, framing our exposition using the language of arbitrary set systems is natural, whereas adopting the constrained language of bicolored graphs would be cumbersome and confusing. 

But in practice, in applications, it is almost always the case that a fixed bicoloring is already given explicitly over all the data (e.g.\ in an author-paper network). Thus in practice, the terms ``bipartite'' and ``bicolored'' graphs are commonly used synonymously; and either the bipartite or hypergraph forms of any particular data set can each be used in different contexts as seen fit.

A bipartite graph with $k$ connected components has $2^k$ possible bicolorings, effectively allowing the edges and vertices to be identified with one or the other of the two parts distinctly within each component. Thus a single bipartite graph may correspond to up to $2^k$ distinct hypergraphs, depending on the bicoloring used. }

In graphs, the structural relationship between two distinct vertices $u$ and $v$ can {\em only} be whether they are adjacent in a {\em single} edge ($\{u,v\} \in E$) or not ($\{u,v\} \not \in E$); and dually, that between two distinct edges $e$ and $f$ can {\em only} be whether they are incident at a single vertex ($e \cap f = \{ v \} \neq \emptyset$) or not ($e \cap f = \emptyset$). 
In hypergraphs, both of these concepts are applicable to {\em sets} of vertices and edges, and additionally become {\em quantitative}. Define $\func{\adj}{2^V}{\ints_{\ge 0}}$ and $\func{\inc}{2^E}{\ints_{\ge 0}}$, in both set notation and (polymorphically) pairwise:
   \[ \adj(U) = | \{ e \supseteq U \} |, \quad
		\adj(u,v) = | \{ e \supseteq \{ u,v \} \} |	\]
	\[ \inc( F ) = | \cap_{e \in F} e |,	\quad
		\inc(e,f)=| e \cap f |	\]
	for $U \subseteq V, u,v \in V, F \subseteq E, e,f \in E$.
	In our example, we have e.g.\ $\adj(a,d)=3, \adj(\{a,c,d\})=1, \inc(1,2)=2, \inc(\{1,2,3\})=1$. These concepts are dual, in that $\adj$ on vertices in $\hg$ maps to $\inc$ on edges in $\hg^*$, and {\it vice versa}. And for singletons, $\adj(\{v\}) = \deg(v) = |e \ni v|$ is the degree of the vertex $v$, while $\inc(\{e\}) = |e|$ is the size of the edge $e$.

\rem{In a hypergraph, adjacency ($|e \supseteq \{u,v\}|$) and incidence ($|e \cap f|$) are quantitative. }

This establishes the basis for extending the central concept of graph theory, a {\bf walk} as a sequential visitation of connected nodes, to hypergraphs. 
Consider a (graph) walk of length $\ell$ as a sequence $W = v_0, e_0, v_1, e_1, \ldots, e_\ell, v_{\ell+1}$ where $v_i,v_{i+1}$ are adjacent in $e_i, 0 \le i \le \ell$, and (dually!) $e_i,e_{i+1}$ are incident on $v_{i+1}, 0 \le i \le \ell-1$. Then $W$ can be equally determined by either the vertex sequence $v_0, \ldots, v_{\ell+1}$, or the edge sequence $e_0, \ldots, e_\ell$. In contrast, with quantitative adjacency and incidence in hypergraphs, sequences of vertices  can be adjacent, and sequences of hyperedges incident, in quantitatively different ways, and need not determine each other. Indeed, vertex sequences become hyperedge sequences in the dual, and {\it vice versa}. For parsimony we work with edgewise walks, and define \cite{AkSJoC19} an {\bf $s$-walk} as a sequence of edges $e_0, e_1, \ldots, e_\ell$ such that $s \le \inc(e_i,e_{i+1})$ for all $0 \le i \le \ell-1$. Thus walks in hypergraphs are characterized not only by length $\ell$, indicating the distance of interaction, but also by ``width'' $s$, indicating a {\em strength} of interaction (see \fig{swalks}).

\mypng{.35}{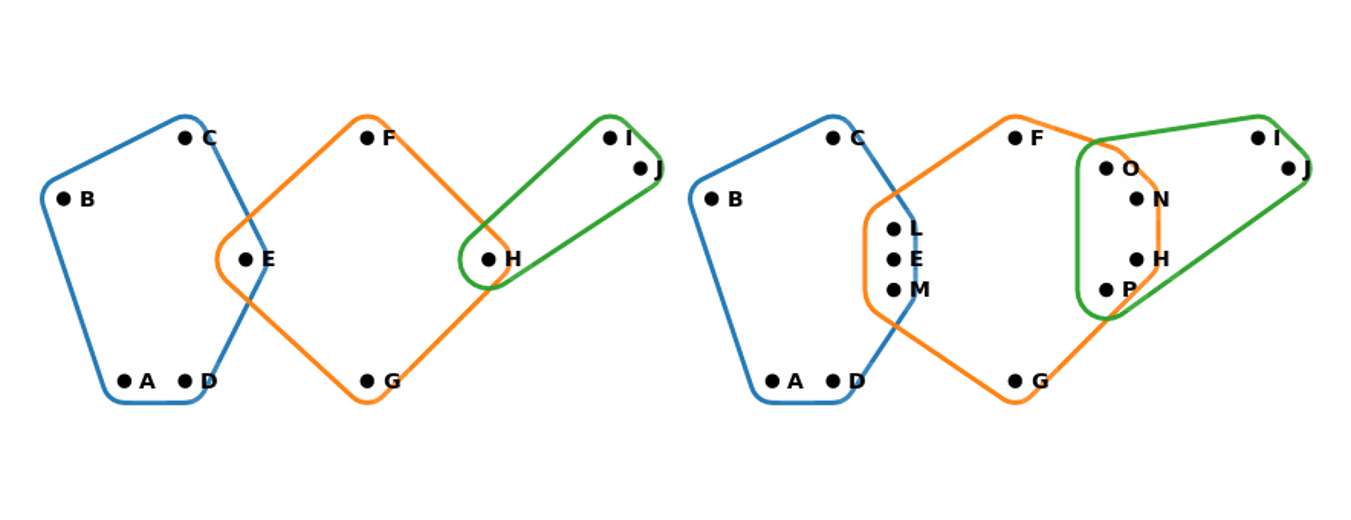}{Two $s$-walks of length $\ell=2$. (Left) Lower width $s=1$. (Right) Higher width $s=3$.}

For a fixed $s > 0$, we define the {\bf $s$-distance} $d_s(e,f)$ between two edges $e,f \in E$ as the length of the shortest $s$-walk between them, or infinite if there is none. Note that a graph walk is a $1$-walk. We have proved  \cite{AkSJoC19} that $s$-distance is a metric, and can thus define the {\bf $s$-diameter} as the maximum $s$-distance between any two edges, and an {\bf $s$-component} as a set of edges all connected pairwise by an $s$-walk. Connected components in graphs are simply 1-components, and our example graph is ``connected'' in that sense, having a single  $1$-connected component. But it has {\em three} 2-components, the hyperedge sets $\{1,2\},\{4\}$, and $\{5\}$. Other network science methods generalize from graphs to hypergraphs \cite{AkSJoC19}, including vertex {\bf $s$-degree}, {\bf $s$-clustering coefficients}, and both {\bf $s$-closeness} and {\bf $s$-betweenness centralities}. 

In graphs, two edges $e,f$ are only incident or not, but in hypergraphs, there could additionally be an {\bf inclusion} relationship $e \sub f$ or $f \sub e$. Indeed, define a {\bf toplex} or {\bf facet} as a maximal edge $e$ such that $\not\! \exists f \supseteq e$, and let $\check{E} \sub E$ be the set of all toplexes. Then the {\bf inclusiveness} $I(\hg) \in [0,1)$ of a hypergraph $\hg$ is the proportion of included edges, that is, the ratio of non-facets to all edges: $I(\hg) \define |E \setminus \check{E}|/|E|$.
For a hypergraph $\hg$, let $\check{\hg} = \tup{V,\check{E}}$ be the {\bf simplification} of $\hg$, and we call $\hg$ {\bf simple} when $\hg = \check{\hg}$. $I(\hg)=0$ iff $\hg = \check{\hg}$ is simple, so that all edges are toplexes. In our example, there are three toplexes $\check{E} = \{ acd, ab, bc \}$, so that $I(\hg)=2/5$.

Maximal $I(\hg)$, on the other hand, is more complicated, and the case when all possible sub-edges are present, so that $E$ is closed by subset. This yields $\hg$ as an {\bf abstract simplicial complex (ASC)}, so that if $e \in E$ and $f \sub e$ then $f \in E$. Let $\widehat{\hg} = \tup{ V, \widehat{E}}$ be the ASC generated by $\hg$, so that $\widehat{E} = \{ g \sub e \}_{e \in E}$ is the closure of the hyperedges by subset. Each hypergraph $\hg$ then maps to a class of hypergraphs we call a {\bf hyperblock} $[\hg]$, so that each pair of hypergraphs $\hg',\hg'' \in [\hg]$ have the same ASC: $\widehat{\hg'}= \widehat{\hg''}$. It follows that they also have the same toplexes: $\widecheck{\hg'}= \widecheck{\hg''}$. 

This results in another representation we call a {\bf simplicial diagram}, shown for our example in \fig{simplicial}. The toplexes $\check{E}$ of $\hg$ are shown as a collection of hyper-tetrahedrons joined where they intersect. This is also sufficient to indicate the ASC $\widehat{\hg}$, and, indeed, all the hypergraphs $\hg' \in [\hg]$ in the hyperblock of $\hg$ are included in the diagram. They are distinguished by additionally labeling the hyperedges (and  circling singletons) actually included in a particular hypergraph $\hg' \in [\hg]$, including both their toplexes and their included edges. In our example, these are $3 = \{d\} \sub 1=\{a,d\} \sub 2=\{a,c,d\}$. Contrast with the singleton $\{b\}$ or graph edge $\{a,c\}$, which are only in the ASC $\hat{\hg}$, and not edges in $\hg$ itself.

\mypng{.25}{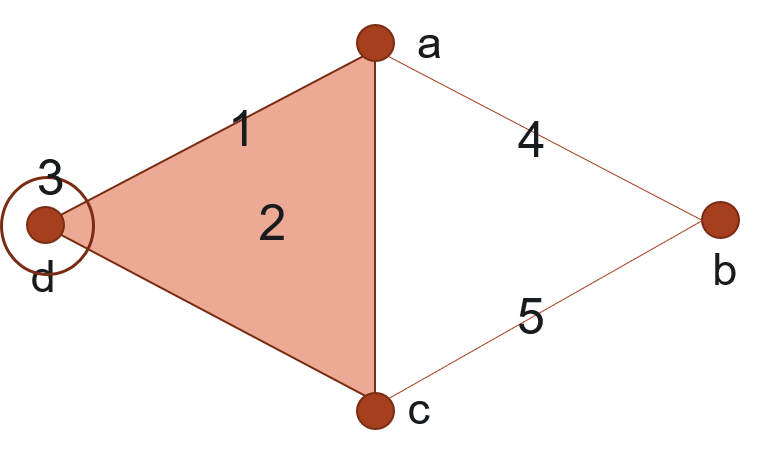}{Example hypergraph $\hg$ as a simplicial diagram.}

\vem
\vem

Given a hypergraph $\hg$, we can define its {\bf $k$-skeleton} $\kskel(\hg) = \{ e \in E \st |e|=k \}$ as the set of hyperedges of size $k$. Each $\kskel(\hg)$ is thus a $k$-uniform sub-hypergraph of $\hg$, and we can conceive of $\hg$ as the disjoint union of its $k$-uniform skeletons: $\hg = \bigsqcup_k \kskel(\hg)$. Where the $k$-skeleton is the set of all edges of size $k$ {\em present} in a hypergraph $\hg$, in contrast the {\bf $k$-section} is the set of all edges of size $k$ {\em implied} by $\hg$, that is, all the vertex sets which are sub-edges of some hyperedge. Formally, $\hg_k = \kskel\left(\widehat{\hg}\right)$, so that the $k$-section is the $k$-skeleton of the ASC of $\hg$, and the ASC is the disjoint union of the sections: $\widehat{\hg} =  \bigsqcup_k  \hg_k$.

\begin{figure}[h]
\centering
\begin{subfigure}[t]{0.15\textwidth}
    \centering
    \includegraphics[width=\textwidth]{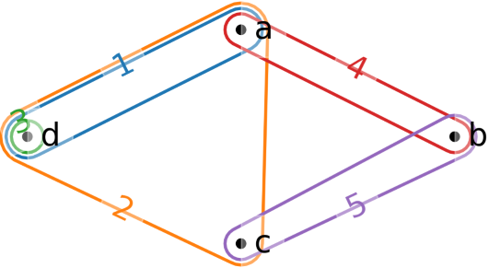}
    \caption{\footnotesize Forward hypergraph $\hg$.}
\end{subfigure}
~
\begin{subfigure}[t]{0.1\textwidth}
    \centering
    \includegraphics[width=\textwidth]{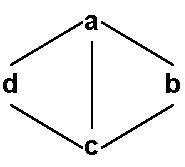}
   	\caption{\footnotesize Forward $2$-section $\hg_2$.}
\end{subfigure}
~
\begin{subfigure}[t]{0.15\textwidth}
    \centering
    \includegraphics[width=\textwidth]{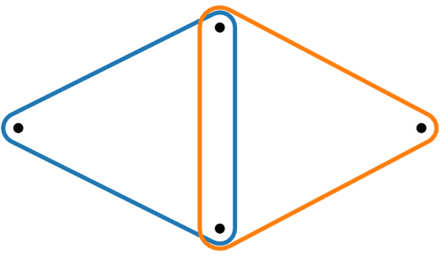}
    \caption{\footnotesize Forward clique reconstruction.}
    \label{fig:biblio_adj}
\end{subfigure}
\\
\begin{subfigure}[t]{0.15\textwidth}
    \centering
    \includegraphics[width=\textwidth]{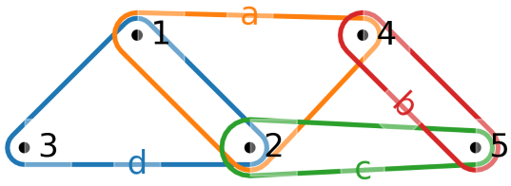}
    \caption{\footnotesize Dual hypergraph $\hg^*$.}
\end{subfigure}
~
\begin{subfigure}[c]{0.15\textwidth}
    \centering
    \includegraphics[width=\textwidth]{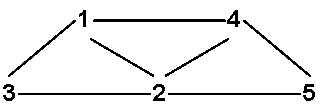}
   	\caption{\footnotesize Line graph $L(\hg) = (\hg^*)_2$.}
\end{subfigure}
~
\begin{subfigure}[c]{0.15\textwidth}
    \centering
    \includegraphics[width=\textwidth]{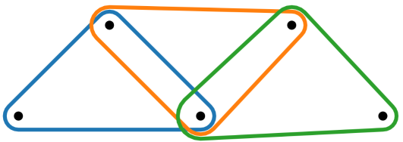}
    \caption{\footnotesize Dual clique reconstruction.}
\end{subfigure}

\caption{\small 2-sections and their clique reconstructions.}

\label{fig:sections}

\end{figure}

Since the $k$-skeletons are all uniform, and any $k$-section or union of $k$-sections is smaller than the entire hypergraph $\hg$, there is substantial interest in understanding  how much information about a hypergraph is available using only them. The $2$-section in particular, which is a graph with adjacency matrix $B B^T$, can be thought of as a kind of ``underlying graph'' of a hypergraph $\hg$. Also of key interest is the 2-section of the dual hypergraph $\hg^*$, called the {\bf line graph} $L(\hg) = (\hg^*)_2$, which dually is a graph with adjacency matrix $B^T B$. As noted above in the discussion of \fig{fig:hgraph}, these are particularly widely used in studies when confronted with complex data naturally presenting as a hypergraph. The limitations of this are evident in the example in \fig{fig:sections}. On the left are our example hypergraph and its dual, and in the center the 2-section $\hg_2$ and the line graph $L(\hg)=(\hg^*)_2$. On the right are the results of taking the maximal cliques of the 2-sections as hyperedges in an attempt to ``reconstruct'' the original hypergraph $\hg$. It is clear how much information is lost.

The ASC $\widehat{\hg}$ is additionally a topological complex, that is, a collection of different $k$-dimensional structures attached together in a particular configuration or pattern. Indeed, the hyperblock $[\hg]$ of a hypergraph $\hg$ generates a number of (finite) topological spaces of interest \cite{EdHHaJ00}. The most cogent of these is the Alexandrov topology with a sub-base consisting of $m$ open sets $T_j = \{ e \supseteq e_j \}_{e \in E}$; that is, for each hyperedge $e_j \in E$, the sub-base element $T_j$ is constructed by collecting all its superedges. The full topology ${\cal T}(\hg)$ is generated by taking all unions of all intersections of these sub-base elements $T_j$. 

The topological space ${\cal T}(\hg)$ will reflect the inherent complexity of the overall ``shape'' of the hypergraph $\hg$. This includes those portions which are connected enough to be contracted, and also the presence of open loops, ``holes'' or ``voids'' of different dimension, which can block such contractions. This is called the {\bf homology} of the space ${\cal T}(\hg)$, and is characterized by its {\bf Betti numbers} $\beta_k, 0 \le k \le \max |e_j|-1$, of $\hg$, indicating the humber of holes of dimension $k$ present in $\hg$. We collect the Betti nubmers to create a {\bf Betti sequence} $\beta = \tup{ \beta_k }_{k=0}^{\max |e_j|-1}$. 

\begin{figure}[h]
\centering
\begin{subfigure}[t]{0.15\textwidth}
    \centering
    \includegraphics[width=\textwidth]{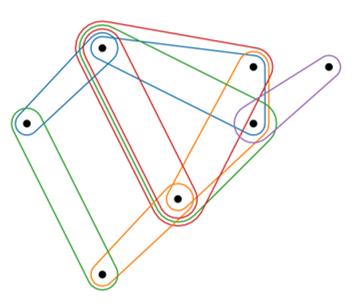}
\end{subfigure}
~
\begin{subfigure}[t]{0.2\textwidth}
    \centering
    \includegraphics[width=\textwidth]{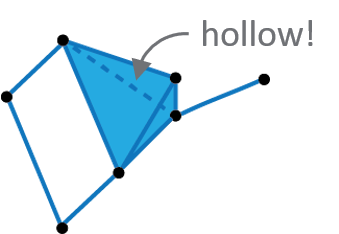}
\end{subfigure}

   	\caption{\small An example simple hypergraph $\check{\hg}$ with $\beta=\tup{1,1,1}$.  (Left) Euler diagram of $\check{\hg}$. (Right) Simplicial diagram of $\protect\widecheck{\hg}$.}

\label{fig:homology}

\end{figure}

The presence of such gaps may invoke questions or hypotheses: what is stopping the connectivity of these holes, of filling them in? In our example, $\beta_1=1$ because of the single open 1-cycle indicated by edges $ab, bc$, and $ca$. Contrast this with the similar cycle $acd$, which is closed in virtue of the hyperedge $2$. $\beta_0=1$, indicating that $\hg$ is 1-connected; while $\beta_2 =0$, so that $\beta=\tup{1,1,0}$. By comparison, the simplicial diagram of the simple hypergraph $\check{\hg}$ shown in \fig{fig:homology} contains a hollow tetrahedron (four triangles surrounding a void), in addition to the open cycle of graph edges on the left. Thus its Betti sequence is $\beta=\tup{1,1,1}$.


\rem{

Furthermore, observe that two vertices belonging to the same set of edges in $H$ correspond to multi-edges in the $H^*$ and isolated vertices in $H$ correspond to empty edges in $H^*$. 
Thus, the generality of our Definition \ref{def:hyp} in permitting multi-edges, empty edges, and isolated vertices ensures the dual of a hypergraph is also a hypergraph. 

}

\section{Example Applications}

Here we illustrate some of the mathematical structures and methods introduced above in brief reports of two example case studies.

\vem

\subsection{Human Gene Set Example}

\vem
While network science using graph theory methods is a dominant discipline in biomolecular modeling, in fact biological systems are replete with many-way interactions likely better represented as hypergraphs. Genes interact in complex combinations, as recorded in a panoply of biomedical databases. We have begun an exploratory examination of the usefulness of hypergraphs in elucidating the relationships between human genes, {\it via} their annotations to semantic categories of human biological processes in the Gene Ontology\footnote{http://geneontology.org/docs/ontology-documentation} and the Reactome pathway database\footnote{https://reactome.org}, chemicals from the Chemical Entities of Biological Interest ontology\footnote{https://www.ebi.ac.uk/chebi/}, and diseases from the Human Disease Ontology\footnote{https://disease-ontology.org/}. These data were selected to help us better understand potential overlaps between gene sets with causative relations in metabolic rare diseases, like phenylketonuria and Alpha-1 antitrypsin deficiency, and their known biological processes and chemical interactions. We also seek to explore potential overlaps in pathway, biological process, and chemical gene sets as a means to elucidate novel gene targets for drug repurposing, which can then be evaluated in the lab. 

\begin{figure}[h]
\centering
\begin{subfigure}[t]{0.45\textwidth}
    \centering
    \includegraphics[width=\textwidth]{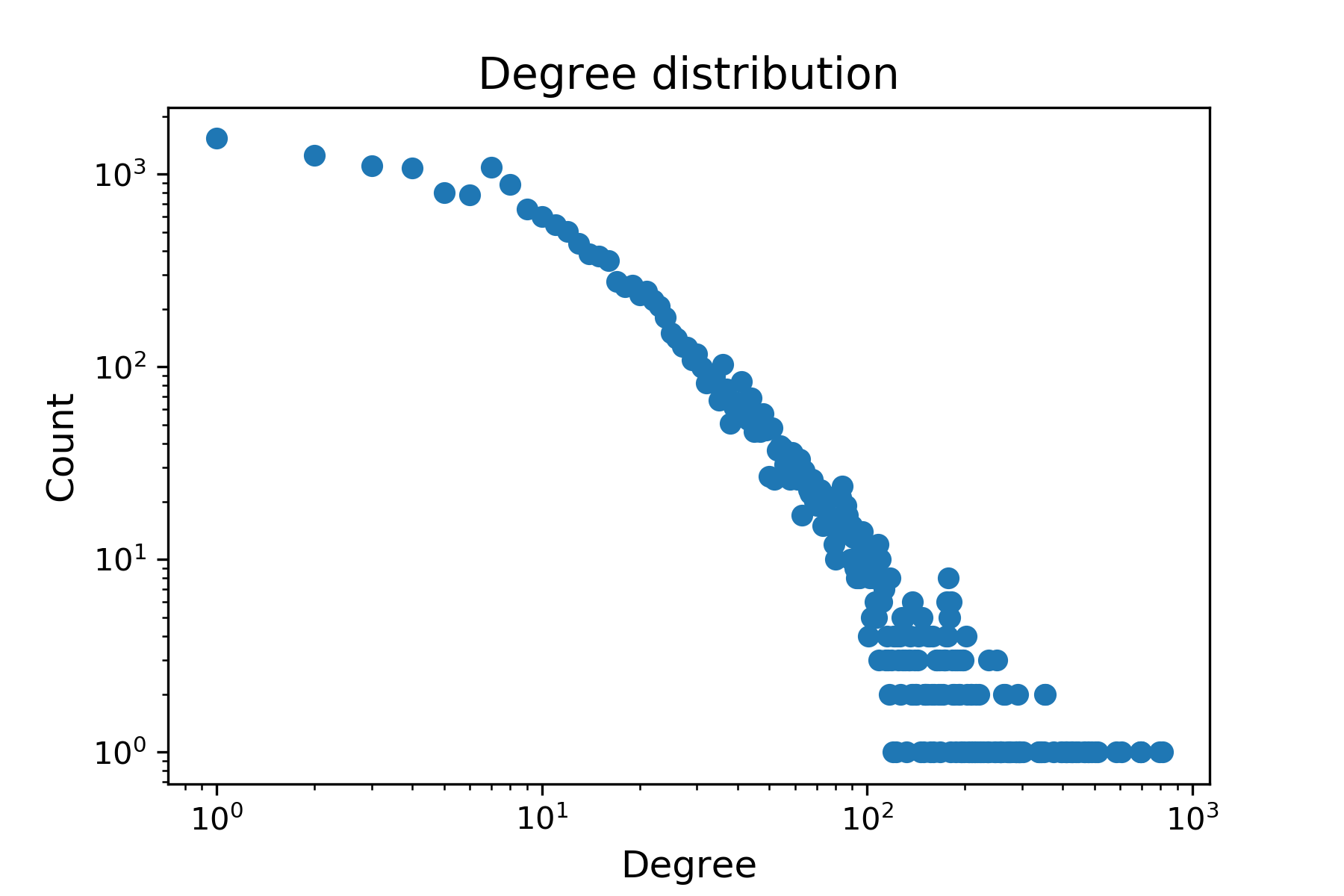}
\end{subfigure}
~
\begin{subfigure}[t]{0.45\textwidth}
    \centering
    \includegraphics[width=\textwidth]{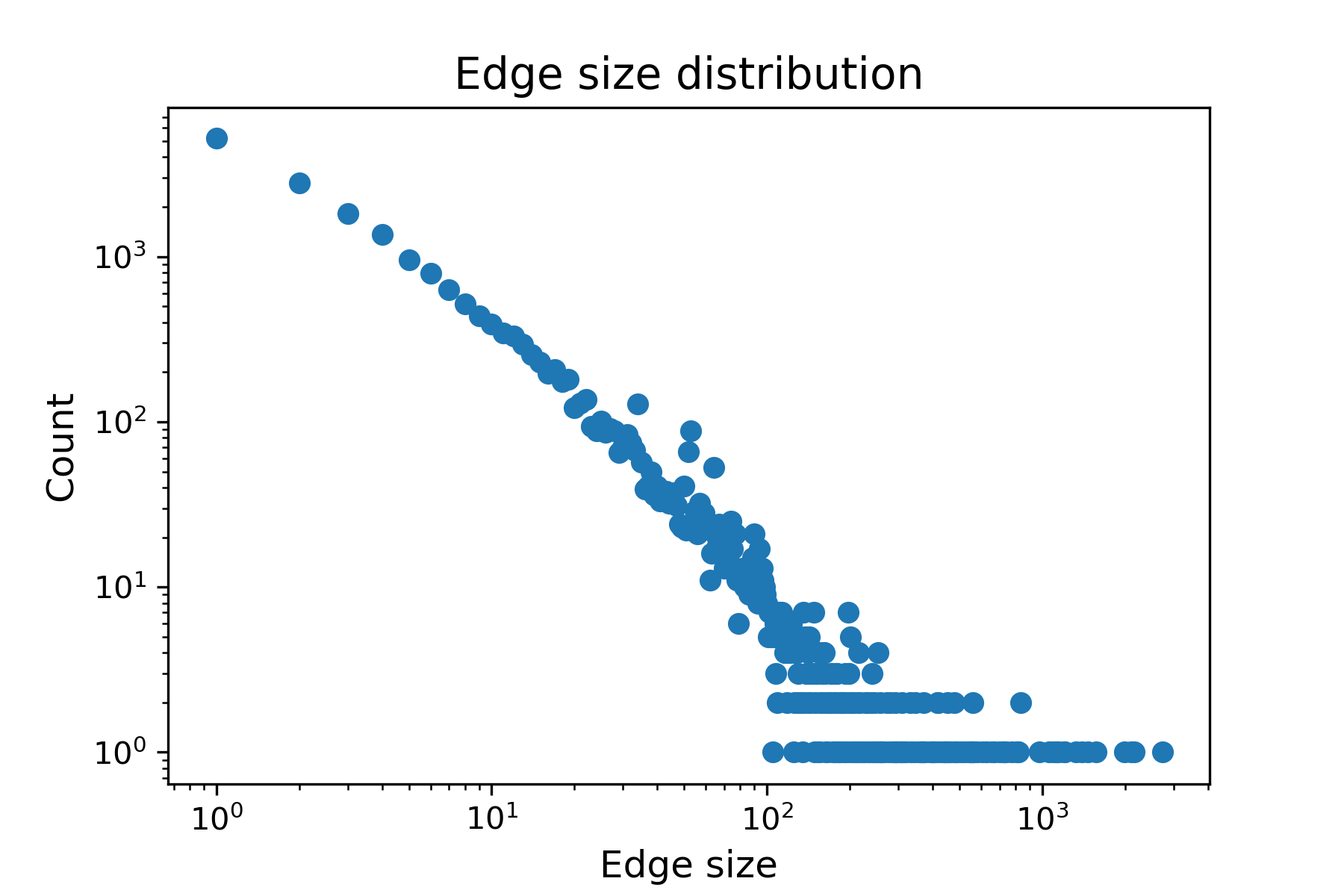}
\end{subfigure}

	\caption{\small Distributions of: (Top) Vertex degree $\deg(v) = \adj(\{v\})$; (Bottom) Edge size $|e|=\inc(\{e\})$.}

\label{dists}

\end{figure}

Data for this analysis were obtained from PheKnowLator v2.0.0.%
	 \footnote{https://github.com/callahantiff/PheKnowLator, downloaded on 03/05/20, see also https://github.com/callahantiff/PheKnowLator/wiki/v2-Data-Sources.}
	 We compiled a hypergraph $\hg = \tup{V,E}$ with $|V|= 17,806$ human genes as vertices against $|E|= 20,568$ annotations as hyperedges, distributed across the source databases as shown in Table~\ref{uc} (noting that there is substantial overlap among these sources). Of these edges, 8,006 are toplexes, yielding an inclusivity of $I(\hg)=61.1\%$, and the density of the incidence matrix $B$ is $0.000926$. \fig{dists} shows the distribution of vertex degree $\deg(v) = \adj(\{v\})$ and edge size $|e|=\inc(\{e\})$, with the expected expoential distribution. 

\begin{table}

\begin{center}

\begin{tabular}{l||r}
{\bf Database} & {\bf Annotations}	\\
\hline
\hline
GO Biological Process	& 12,305	\\
Reactome Pathways	& 2,291	\\
Chemicals	& 3,289	\\
Diseases	& 2,683	\\
\hline
{\bf Total}	& {\bf 20,568}	\\

\end{tabular}

\caption{Distribution of annotations across biological databases.}

\label{uc}

\end{center}

\end{table}


\begin{figure}[h]
\centering
\begin{subfigure}[t]{0.45\textwidth}
    \centering
    \includegraphics[width=\textwidth]{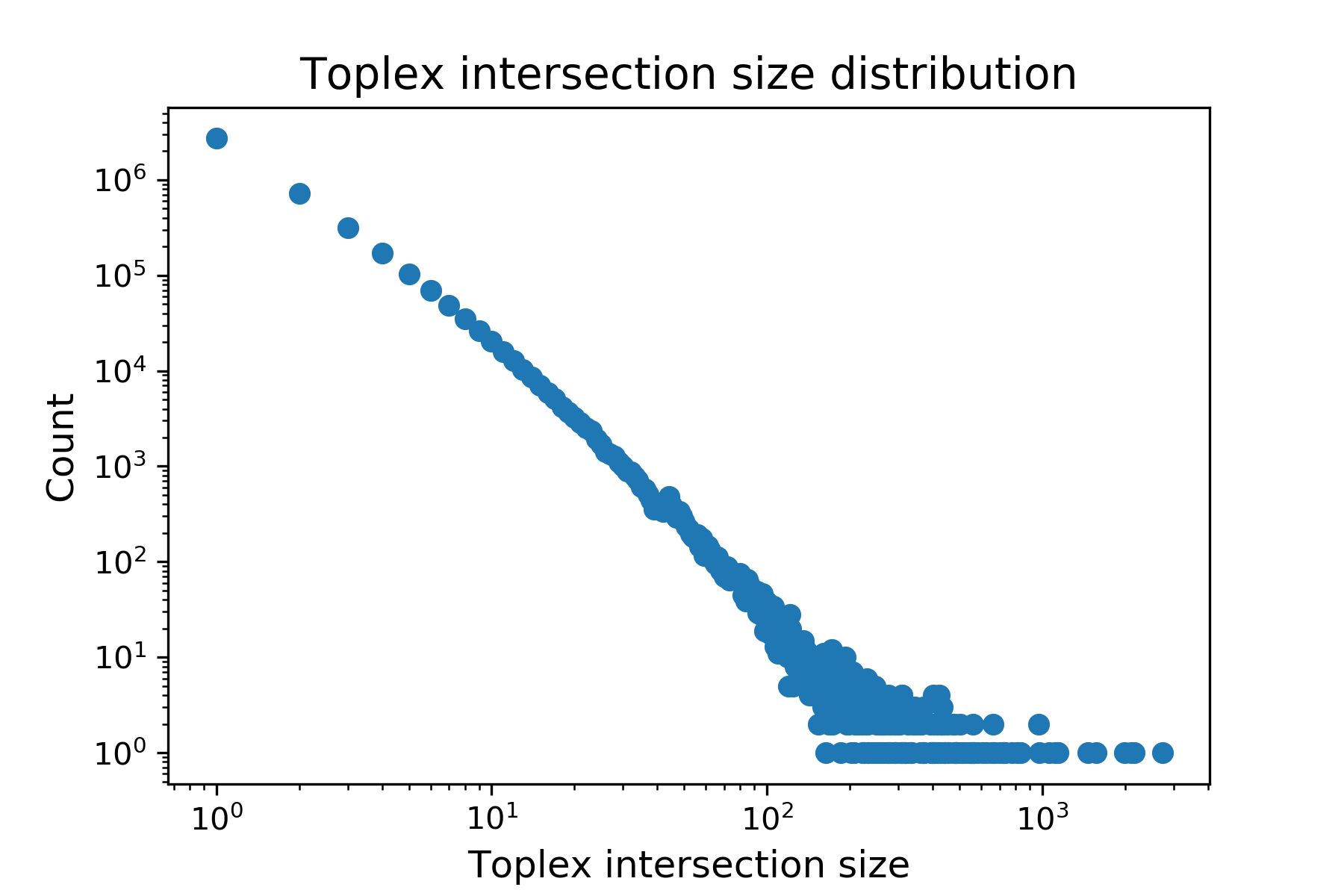}
\end{subfigure}
~
\begin{subfigure}[t]{0.45\textwidth}
    \centering
    \includegraphics[width=\textwidth]{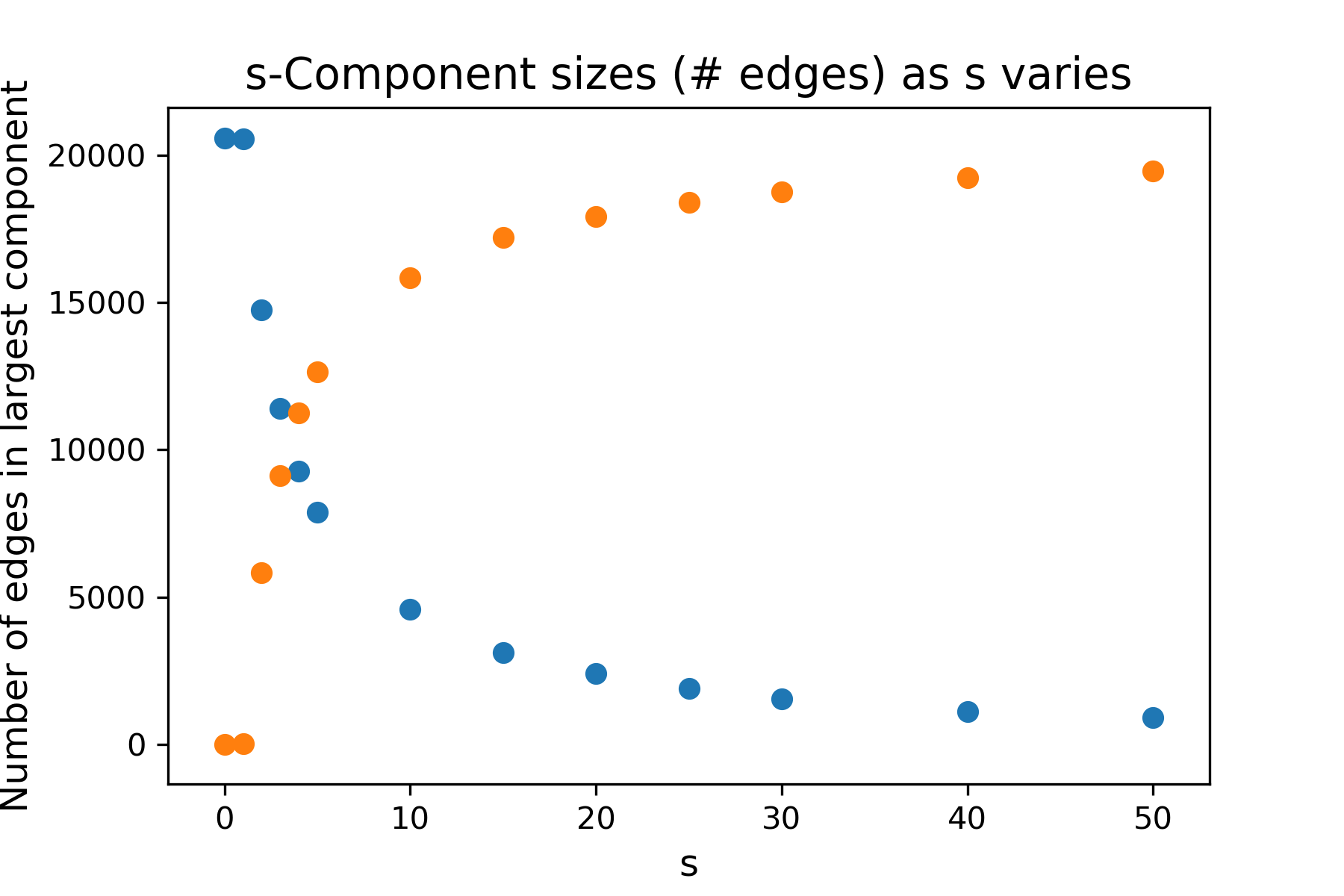}
\end{subfigure}

	\caption{\small (Top) Distribution of the size of toplex intersections: $\inc(e,f)$ for $e,f \in \check{E}$. (Bottom) \# components (orange) and size of largest component (blue).}

\rem{

	\caption{\small (Left) \# components (orange) and size of largest component (blue). (Right) Distributional statistics for the component size distribution as a function of width $s$, from $s=0$ (lower left) to $s=50$ upper right.}

}

\label{comps}

\end{figure}

\fig{dists} shows only the lowest, ``first order'' distribution of the hypergraph structure, the adjacency and incidence of singletons. Consideration of higher-order interactions would require expensive combinatorial calculations of, for example, $k$-way intersections and hyperedge inclusions of arbitrary sets of vertices. A modest step towards that goal in our methodology is first to focus on toplexes, which determine the topological structure, and then their pairwise intersections: $\inc(e,f)$ for $e,f \in \check{E}$. This is shown in the top of \fig{comps}, which reveals a long tail, indicating a significant number of pairs of annotations with large intersections of genes. Attending to incidences of even higher order would reveal the increasingly rich complex interactions of gene sets.

\rem{

12,305 biological processes, 2,291 pathways, 3,289 chemicals, and 2,683 diseases. The total number of unique genes for each gene-biological entity set ranged widely with largest total number of unique genes observed for biological processes (17,252 genes, avg=11.198, min=1, max=1,133), followed by pathways (10,677 genes, avg=48.134, min=1, max=2,732), chemicals (6,445 genes, avg=14.844, min=1, max=1,567), and diseases (5,203 genes, avg=15.750, min=1, max=5,203).

}


Of even more interest is bottom of \fig{comps}, which shows distribution information about the connected $s$-components for different intersection levels $s$. On the top the number of $s$-components is shown in orange, and the size of the largest component in blue, all as a function of increasing $s$. Expectedly these appear monotonic increasing and decreasing respectively, but it's notable that even for large $s$ there persist some very large components, again demonstrating the large multi-way interactions amongst these gene sets.

\vem

\subsection{DNS Cyber Example}

The Domain Name System (DNS) provides a decentralized service to
map from domain names (e.g.,
\code{www.google.com}) to IP addresses. Perhaps somewhat counter-intuitively, DNS data present
naturally as a hypergraph, in being a many-many relationship
between domains and IPs. While typically this relationship is
one-to-one, with each domain uniquely identifying a single IP address
and {\it vice versa}, there are a number of circumstances which can
violate this, for example domain aliasing,  hosting services where  one IP serves multiple  websites, or duplicated IPs to manage loads against popular domains. 

ActiveDNS (ADNS) is a data set maintained by the Astrolavos Lab at Georgia
Institute of Technology.\footnote{https://activednsproject.org} It submits daily DNS lookups for popular zones (e.g., .com, .net, .org) and lists of domain names.
Using data from April 26, 2018 as an example, 
this day consists of 1,200 Avro files with each file containing on average 900K records.
Our hypergraph representation coded each
domain (hyperedges $e \in E$) as a collection of its IPs (vertices $v \in V$). A small portion of the incidence matrix $B$ is shown in \fig{adns_matrix}.

\mypng{.4}{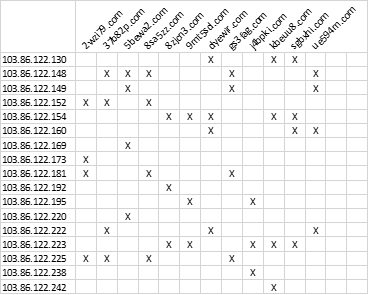}{Portion of the incidence matrix for ADNS data.}

To identify some of the simplest hypergraph-specific properties, we looked \cite{JoCAkS20a} specifically at the 2-components, and identified the one with maximum 2-diameter (6), which is
shown in Figure \ref{flux}. 
The IP addresses in this component all belong to the IP range 103.86.122.0/24 and the domains are registered to GMO INTERNET, INC according to WHOIS records.
Moreover, current DNS queries for most of these domains at a later date resolve to IPs in the range 103.86.123.0/24 and have a ``time to live'' of only 120 seconds.
This pattern of quickly changing of IP address is consistent with the ``fast flux'' DNS technique which can be used by botnets to hide malicious content delivery sites and make networks of malware more difficult to discover \cite{fastflux}.

\mypng{.35}{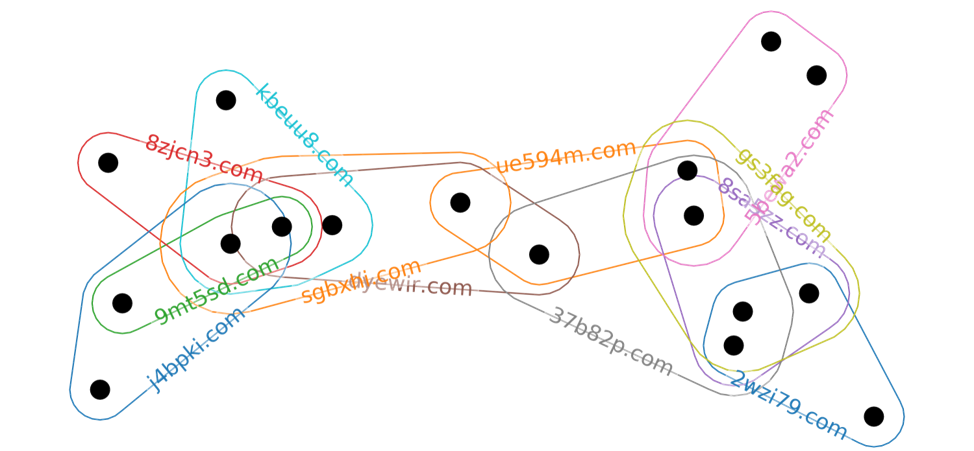}{The 2-component with largest 2-diameter, possibly indicating fast flux behavior.}

Other 2-components reveal non-trivial homologies, three of which are shown in \fig{3dnsgraphs}. DNS1 has $\beta=\tup{1,1,0,0,\ldots}$, with a visible hole surrounded by a $1$ dimensional loop on the top. DNS2 has $\beta=\tup{1,1,2,0,\ldots}$, and DNS3 has $\beta=\tup{1,3,1,0,\ldots}$, indicating one and three 1-dimensional holes, and two and one 2-dimensional voids, respectively. The open loops in DNS2 and DNS3 are harder to visualize, so \fig{tetra_dns1} shows a simplicial diagram of one of the two 2-dimensional voids in DNS2. There are two solid tetrahedrons for the domains potterybarn.com and pbteen.com, each with four IPs, three of which (those ending in .160, .9, and .105) they share. Then wshome.com is a triangle in the foreground, and westelm.com a triangle behind (see caption for details). These are effectively ``transparent window panes'' surrounding a hollow tetrahedral space. Identification of such multi-dimensional open-loop structures affords the opportunity to consider these as hypotheses: on what basis is the Pottery Barn company structuring its multiple domains over their multiple IPs in this complex pattern? 

\rem{

dns1=the one with bf7b8-q.com

i'm pretty sure dns2= the one with pbteen.com 

leaving prommersberger.com'  as the odd man out
 
}

\mypng{.3}{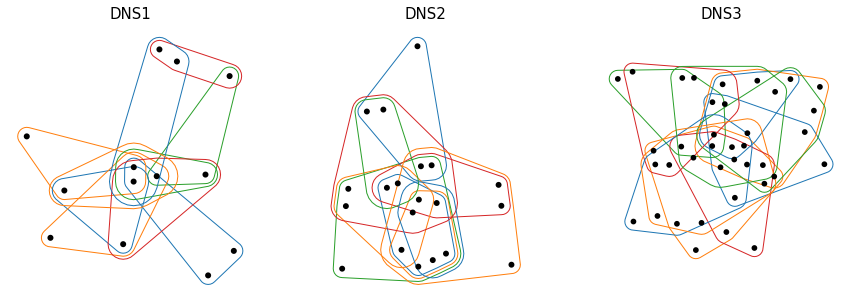}{Three 2-components with non-trivial homologies.}

\mypng{.5}{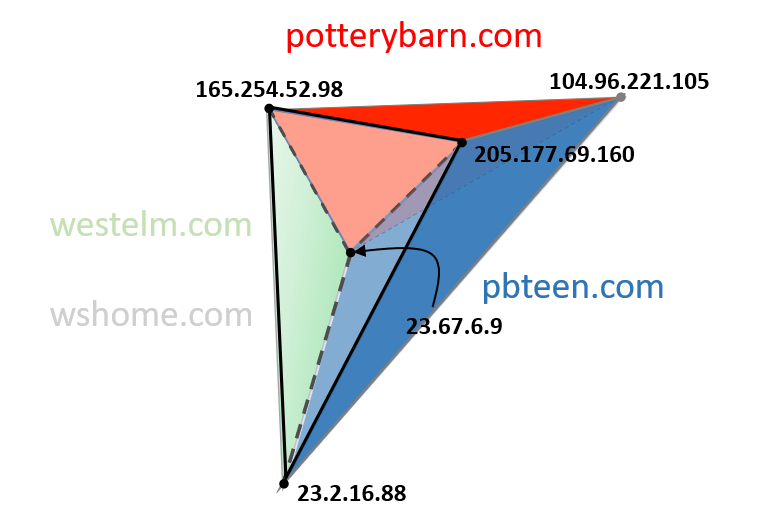}{Simplicial diagram of the complex pattern of IPs shared by some Pottery Barn domains around one of the two 2-dimensional voids in DNS2 of \fig{3dnsgraphs}. wshome.com is the transparent foreground triangle with IPs ending in .160, .88 and .98; and westelm.com the background triangle with IPs ending in .98, .88, and .9. Potterybarn.com and pbteen.com are solid tetrahedrons with four IPs each.}

\rem{

\red{Brett: (Suggestion) in conclusion a small table of measures unique to hypergraphs. No more than .25 of a page to give flavor of capabilities}

}

\vem
\vem
\vem
\vhalf

\small

\bibliographystyle{plain}
\bibliography{hypergraphs_ismb,HICSS_2019_eap_bib,sWalkRefs,iccs_20}

\begin{thebibliography}{10}

\bibitem{AkSJoC19}
Sinan~G Aksoy, Cliff~A Joslyn, Carlos~O Marrero, B~Praggastis, and Emilie~AH
  Purvine.
\newblock Hypernetwork science via high-order hypergraph walks.
\newblock 2019.
\newblock submitted, https://arxiv.org/abs/1906.11295.

\bibitem{alon1990transversal}
Noga Alon.
\newblock Transversal numbers of uniform hypergraphs.
\newblock {\em Graphs and Combinatorics}, 6(1):1--4, 1990.

\bibitem{AuGFaP01}
Giorgio Ausiello, Paolo~G Fanciosa, and Daniele Frigioni.
\newblock Directed hypergraphs: Problems, algorithmic results, and a novel
  decremental approach.
\newblock In {\em ICTCS 2001, LNCS}, volume 2202, pages 312--328. 2001.

\bibitem{Barabasi2016}
Albert~László Barabási.
\newblock {\em Network Science}.
\newblock Cambridge University Press, Cambridge, 2016.

\bibitem{Barber2007}
Michael~J. Barber.
\newblock Modularity and community detection in bipartite networks.
\newblock {\em Physical Review E}, 76(6), dec 2007.

\bibitem{berge1973graphs}
Claude Berge and Edward Minieka.
\newblock {\em {Graphs and Hypergraphs}}.
\newblock North-Holland, 1973.

\bibitem{chung1993laplacian}
Fan Chung.
\newblock The laplacian of a hypergraph.
\newblock {\em Expanding graphs (DIMACS series)}, pages 21--36, 1993.

\bibitem{cooper2012spectra}
Joshua Cooper and Aaron Dutle.
\newblock Spectra of uniform hypergraphs.
\newblock {\em Linear Algebra and its Applications}, 436(9):3268--3292, 2012.

\bibitem{Devine2006}
K.D. Devine, E.G. Boman, R.T. Heaphy, R.H. Bisseling, and U.V. Catalyurek.
\newblock Parallel hypergraph partitioning for scientific computing.
\newblock In {\em Proceedings 20th {IEEE} International Parallel {\&}
  Distributed Processing Symposium}. {IEEE}, 2006.

\bibitem{Dewar2018}
Megan Dewar, John Healy, Xavier P{\'{e}}rez-Gim{\'{e}}nez, Pawe{\l} Pra{\l}at,
  John Proos, Benjamin Reiniger, and Kirill Ternovsky.
\newblock Subhypergraphs in non-uniform random hypergraphs.
\newblock {\em Internet Mathematics}, mar 2018.

\bibitem{dinur2005hardness}
Irit Dinur, Oded Regev, and Clifford Smyth.
\newblock The hardness of 3-uniform hypergraph coloring.
\newblock {\em Combinatorica}, 25(5):519--535, 2005.

\bibitem{Dorfler1980}
W.~D\"{o}rfler and D.~A. Waller.
\newblock A category-theoretical approach to hypergraphs.
\newblock {\em Archiv der Mathematik}, 34(1):185--192, dec 1980.

\bibitem{EdHHaJ00}
Herbert Edelsbrunner and John~L Harer.
\newblock {\em Computational Topology: An Introduction}.
\newblock AMS, 2000.

\bibitem{Estrada2006}
Ernesto Estrada and Juan~A. Rodr{\'{\i}}guez-Vel{\'{a}}zquez.
\newblock Subgraph centrality and clustering in complex hyper-networks.
\newblock {\em Physica A: Statistical Mechanics and its Applications},
  364:581--594, may 2006.

\bibitem{Fong2018}
Brendan Fong and David~I Spivak.
\newblock Hypergraph categories.
\newblock 2019.

\bibitem{GaGLoG93}
Giorgio Gallo, Giustino Longo, and Stefano Pallottino.
\newblock Directed hypergraphs and applications.
\newblock {\em Discrete Applied Mathematics}, 42:177--201, 1993.

\bibitem{IaIPeG19}
I~Iacopini, G~Petri, A~Barrat, and V~Latora.
\newblock Simplicial models of social contagion.
\newblock {\em Nature Communications}, 10:2485, 2019.

\bibitem{fastflux}
jamie.riden.
\newblock {How Fast-Flux Service Networks Work}.
\newblock \url{http://www.honeynet.org/node/132}.
\newblock Accessed: 2018-11-26.

\bibitem{JaMLuL18}
M~A Javidian, L~Lu, M~Valtorta, and Z~Qang.
\newblock On a hypergraph probabilistic graphical model, 2018.

\bibitem{JoJ13}
Jeffrey Johnson.
\newblock {\em Hypernetworks in the Science of Complex Systems}.
\newblock Imperial College Press, London, 2013.

\bibitem{JoCAkS20a}
Cliff~A Joslyn, Sinan Aksoy, Dustin Arendt, J~Firoz, Louis Jenkins, Brenda
  Praggastis, Emilie~AH Purvine, and Marcin Zalewski.
\newblock Hypergraph analytics of domain name system relationships.
\newblock In {\em 17th Wshop. on Algorithms and Models for the Web Graph (WAW
  2020), Lecture Notes in Computer Science}, 2020.
\newblock in press.

\bibitem{Karypis2000}
George Karypis and Vipin Kumar.
\newblock Multilevel k-way hypergraph partitioning.
\newblock {\em {VLSI} Design}, 11(3):285--300, jan 2000.

\bibitem{Kirkland2017}
Steve Kirkland.
\newblock Two-mode networks exhibiting data loss.
\newblock {\em Journal of Complex Networks}, 6(2):297--316, aug 2017.

\bibitem{klamt2009hypergraphs}
S~Klamt, U-U Haus, and F~Theis.
\newblock Hypergraphs and cellular networks.
\newblock {\em PLoS Computational Biology}, 5(5):e1000385, 2009.

\bibitem{krivelevich2003approximate}
Michael Krivelevich and Benny Sudakov.
\newblock Approximate coloring of uniform hypergraphs.
\newblock {\em Journal of Algorithms}, 49(1):2--12, 2003.

\bibitem{Larremore2014}
DB Larremore, A Clauset, and AZ Jacobs.
\newblock Efficiently inferring community structure in bipartite networks.
\newblock {\em Physical Review E}, 90(1), jul 2014.

\bibitem{Latapy2008}
Matthieu Latapy, Cl{\'{e}}mence Magnien, and Nathalie~Del Vecchio.
\newblock Basic notions for the analysis of large two-mode networks.
\newblock {\em Social Networks}, 30(1):31--48, jan 2008.

\bibitem{LeWReG19}
W~Leal and G~Restrepo.
\newblock Formal structure of periodic system of elements.
\newblock {\em Proc. R. Soc. A.}, 475, 2019.

\bibitem{MiM00}
M~Minas.
\newblock Hypergraphs as a unifrom diagram representation model.
\newblock In {\em Proc. 6th Int. Workshop on Theory and Applications of Graph
  Transformations}, 2000.

\bibitem{PaAPeG17}
A~Patania, G~Petri, and F~Vaccarino.
\newblock Shape of collaborations.
\newblock {\em EPJ Data Science}, 6:18, 2017.

\bibitem{Robins2004}
Garry Robins and Malcolm Alexander.
\newblock Small worlds among interlocking directors: Network structure and
  distance in bipartite graphs.
\newblock {\em Computational {\&} Mathematical Organization Theory},
  10(1):69--94, may 2004.

\bibitem{rodl2004regularity}
Vojt{\v{e}}ch R{\"o}dl and Jozef Skokan.
\newblock Regularity lemma for k-uniform hypergraphs.
\newblock {\em Random Structures \& Algorithms}, 25(1):1--42, 2004.

\bibitem{Schmidt2019}
Martin Schmidt.
\newblock Functorial approach to graph and hypergraph theory.
\newblock 2019.

\end{thebibliography}


\end{document}